%% file: main.tex
\documentclass{article}

\usepackage[preprint]{neurips_2025}

\usepackage[utf8]{inputenc} 
\usepackage[T1]{fontenc}    
\usepackage{hyperref}       
\usepackage{url}            
\usepackage{booktabs}       
\usepackage{amsfonts}       
\usepackage{nicefrac}       
\usepackage{microtype}      
\usepackage{xcolor}         

\usepackage{amsmath}
\usepackage{subcaption}
\usepackage{todonotes}
\usepackage{algorithm}
\usepackage{algorithmic}
\usepackage{tikz}

\graphicspath{{figures/}}

\newif\ifPaperDraft
\PaperDraftfalse 
\input{util}

\title{nvPAX: Constrained Optimization for Dynamic Power Allocation in Hierarchical and Multi-Tenant Systems}

\author{%
  Hadar Sivan \\
  NVIDIA \\
  \texttt{hsivan@nvidia.com} \\
  \And
  Gil Shabat \\
  NVIDIA \\
  \texttt{gshabat@nvidia.com} \\
  \And
  Yoel Shkolnisky \\
  Tel Aviv University \& NVIDIA \\
  \texttt{yoelsh@tauex.tau.ac.il} \\
}

\begin{document}

\maketitle

\input{sections/abstract.tex}
\input{sections/background.tex}
\input{sections/related_work.tex}
\input{sections/policy_requirements.tex}
\input{sections/algorithmic_framework.tex}
\input{sections/experiments.tex}
\input{sections/conclusion.tex}

\bibliographystyle{plain}
\bibliography{references}

\input{sections/appendix.tex}

\end{document}

%% file: util.tex
\makeatletter
\@ifundefined{ifPaperDraft}{%
  \newif\ifPaperDraft
  \PaperDrafttrue
}{}
\makeatother

\ifPaperDraft
  \usepackage[draft,markup=underlined]{changes}
\else
  \usepackage[final]{changes}
\fi

\definechangesauthor[name={HS}, color=blue]{hs}
\definechangesauthor[name={GS}, color=purple]{gs}
\definechangesauthor[name={YS}, color=red]{ys}
\definechangesauthor[name={Reviewer}, color=red]{rev}

\makeatletter
\@ifpackageloaded{todonotes}{%
  \setlength{\marginparwidth}{2.6cm}

}{%

}
\makeatother

%% file: sections/abstract.tex
\begin{abstract}
Power oversubscription is increasingly central to datacenter operation as power density grows, making it necessary to dynamically allocate limited power budgets across devices based on real-time demand.
Existing approaches typically assume flat power domains, whereas in practice power distribution is hierarchical and allocation decisions must additionally respect tenant-level contractual constraints.
We present nvPAX, a constrained-optimization policy that computes feasible power allocations at every control step via a three-phase hybrid QP/LP procedure. Phase~I allocates power with minimum deviation from each device's power request, while respecting job priorities. Phase~II fairly distributes excess power among active devices. Phase~III fairly distributes any remaining power to idle devices. The rationale behind the three phases is to allow power oversubscription while maximizing datacenter utilization.
On a trace-driven large-scale simulation using GPU power telemetry from a production datacenter, nvPAX runs with a mean wall-clock time of 264.69~ms per allocation interval and achieves a mean satisfaction ratio of 98.92\%, outperforming static equal-share allocation and providing robustness beyond greedy proportional allocation in the presence of non-uniform hierarchical bottlenecks.

\end{abstract}

%% file: sections/background.tex
\section{Introduction}\label{the-problem}

Current and future datacenters are designed and built with power oversubscription: the aggregate peak power consumption of devices exceeds the power available from the utility feed or from intermediate distribution nodes~\mbox{\cite{patel2024characterizing}}.
Power oversubscription can increase effective cluster throughput by exploiting temporal diversity in demand, since not all jobs draw peak power simultaneously, so provisioning for the sum of per-device peaks is inefficient.
However, oversubscription increases the risk of violating upstream power distribution network (PDN) constraints (e.g., rack, power distribution unit (PDU), or facility limits), which can trigger capping actions or outages.
This motivates fast control policies that, at every decision point, enforce feasible power allocations close to the per-device power requests.
Operators therefore require dynamic allocation policies that translate per-device requests into enforceable power limits while respecting PDN constraints and tenant-level contractual guarantees.

A variety of approaches exist in the literature and in practice.
Control-theoretic methods enforce power limits at multiple levels of the hierarchy (e.g., SHIP~\cite{ship2009} at rack, PDU, and datacenter) but typically do not incorporate tenant-level SLAs or explicit fairness across devices.
Linear programming has been used for energy-aware server provisioning and cluster-level performance~\cite{guenter11managing}, though without modeling a full hierarchical oversubscribed PDN or tenant domains.
Prediction- and learning-based systems improve oversubscription or service level objective (SLO) under capping~\cite{kumbhare21atc,sakalkar20asplos,parm23neurips} but do not guarantee that every control decision satisfies all physical and contractual constraints. Reinforcement learning (RL) and related methods~\cite{rldatacenter2025} generally lack deterministic feasibility at each step.
Commercial solutions and disclosed patents often target a single level (e.g., rack or server) load monitoring and provisioning~\cite{dell2024fuzzy}, rather than per-device allocation under a full PDN tree and tenant SLAs.
Section~\ref{related-work} reviews and compares these approaches.
A common pattern in production datacenters is a closed-loop power capping system that periodically collects power data, predicts demand, applies a policy to set power caps, and configures devices accordingly.
PRS is one such framework~\mbox{\cite{sivan2024prs}}; it currently supports only a flat (non-hierarchical) power domain.
The gap we address is that no existing method jointly handles a full hierarchical PDN and tenant SLA constraints, while providing a provably fair, optimal policy in a single formulation suitable for fast control loops.

\paragraph{Hierarchical PDN and Physical Constraints.}
The actual PDN in datacenters is hierarchical: a tree from the utility feed down to devices.
At each level, oversubscription can occur. Specifically, the utility feed supplies a finite budget to the datacenter, and each data hall's PDU is allocated a share such that the sum across halls may exceed the utility budget. This structure continues through rack PDUs down to devices, where the combined nominal power of devices may exceed the capacity of their parent node.
Power consumption at any level must not exceed the capacity of its parent.
These hierarchical capacity constraints are vertical in that they apply to each subtree along the PDN tree.
In addition, each device has minimum and maximum permissible power limits, and its allocated power must lie within this range.
We refer to these as physical constraints.

\paragraph{Tenant SLAs.}
In cloud environments, the provider assigns servers to tenants; each tenant domain is governed by a service level agreement (SLA) that may include, for example, a minimum power allocation guarantee over the tenant's domain, per-device minimum guarantees, domains with ``max-Q'' or ``max-P'' constraints (GPU power profiles that cap a device at a reduced or full thermal design power, respectively)~\cite{narayanaswamy2025datacenterenergyoptimizedpower}, and the option for tenants to opt in or out of dynamic power allocation.
These form a set of service-level constraints that any power allocation must satisfy.
Unlike the vertical PDN constraints, tenant SLAs are horizontal, namely, they can couple devices across different branches of the physical hierarchy.

\paragraph{Priorities and Active Versus Idle Devices.}
Power allocation should respect job priorities, where higher-priority jobs receive precedence when fulfilling requests. It should also distinguish active devices (assigned to running jobs) from idle ones, prioritizing active devices and allocating power to idle devices only when headroom remains.

\paragraph{Fairness.}
In distributed multi-device workloads (e.g., multi-GPU training), end-to-end progress is often determined by the slowest device, so uneven allocations among devices can create stragglers and bottlenecks even when aggregate allocated power is high.
A practical policy should therefore distribute shortages and surplus so as to avoid disproportionate degradation of any single device within a job or priority class.

\begin{figure}[t]
\centering
\begin{tikzpicture}[
	scale=0.86,
	level 1/.style={sibling distance=5.0cm, level distance=1.45cm},
	level 2/.style={sibling distance=2.35cm, level distance=1.35cm},
	level 3/.style={sibling distance=1.45cm, level distance=1.22cm},
	level 4/.style={sibling distance=0.8cm, level distance=1.05cm},
	edge from parent/.style={draw, -latex},
	every node/.style={draw, rounded corners=2pt, font=\small, align=center, inner sep=3pt},
	power/.style={minimum width=1.4cm, minimum height=0.55cm},
	hall/.style={minimum width=1.3cm, minimum height=0.5cm},
	rack/.style={minimum width=1.1cm, minimum height=0.45cm},
	device/.style={font=\scriptsize, minimum width=0.55cm, minimum height=0.32cm},
	tenantA/.style={fill=blue!22},
	tenantB/.style={fill=orange!22},
	tenantC/.style={fill=green!22},
	grow=down,
]
\node[power] (utility) at (1.9,0) {Utility\\10 MW}
	child { node[hall] (hall1) {Hall 1\\6 MW}
		child { node[rack] (rack1) {Rack 1\\4 MW}
			child { node[device, tenantA] (S1) {S1} }
			child { node[device, tenantA] (S2) {S2} }
		}
		child { node[rack] (rack2) {Rack 2\\4 MW}
			child { node[device, tenantB] (S3) {S3} }
			child { node[device, tenantC] (S4) {S4} }
		}
	}
	child { node[hall] (hall2) {Hall 2\\6 MW}
		child { node[rack] (rack3) {Rack 3\\4 MW}
			child { node[device, tenantC] (S5) {S5} }
			child { node[device, tenantC] (S6) {S6} }
		}
		child { node[rack] (rack4) {Rack 4\\4 MW}
			child { node[device, tenantA] (S7) {S7} }
			child { node[device, tenantB] (S8) {S8} }
		}
	};

\draw[
	dashed,
	rounded corners=6pt,
	line join=round,
	line cap=round,
	thick,
	draw=green!50!black
]
	([xshift=-1.5mm,yshift=3mm]S4.north west)
	rectangle
	([xshift=1.5mm,yshift=-4mm]S6.south east);

\draw[
	dash pattern=on 3.2pt off 3.8pt on 0.8pt off 3.8pt,
	rounded corners=35pt,
	line join=round,
	line cap=round,
	thick,
	draw=black!70
]
	([yshift=12mm]hall2.north)
	-- ([xshift=-10mm,yshift=-2mm]S5.south west)
	-- ([xshift=10mm,yshift=-2mm]S8.south east)
	-- cycle;

\node[draw=none, anchor=east, font=\scriptsize, align=left] at ([xshift=50mm,yshift=-2mm]utility.east) {
\begin{tabular}{@{}l@{}}
\textbf{Tenants:}\\
\tikz\draw[draw,fill=blue!22] (0,0) rectangle (0.22,0.22);~Tenant A\\
\tikz\draw[draw,fill=orange!22] (0,0) rectangle (0.22,0.22);~Tenant B\\
\tikz\draw[draw,fill=green!22] (0,0) rectangle (0.22,0.22);~Tenant C\\
\end{tabular}
};

\path (utility.south) -- (hall1.north) coordinate[midway] (mid_utility_hall);
\path (hall1.south) -- (rack1.north) coordinate[midway] (mid_hall_rack);
\node[draw=none, anchor=east, font=\scriptsize] at ([xshift=-25mm,yshift=-10]mid_utility_hall -| utility.west)
  {Halls: 6~MW $\times$ 2 = \\12~MW \;(1.2$\times$)};
\node[draw=none, anchor=east, font=\scriptsize] at ([xshift=-36mm,yshift=-10]mid_hall_rack -| utility.west)
  {Racks: 4~MW $\times$ 4 =\\ 16~MW \;(1.6$\times$)};
\end{tikzpicture}
\caption{Datacenter power and tenant layout illustrating hierarchical oversubscription.
The tree represents the physical PDN from utility down to servers.
Each node is labeled with its power capacity; notably, the sum of children's capacities exceeds the parent's capacity at multiple levels (e.g., 12~MW aggregate hall capacity vs.\ 10~MW utility feed), necessitating dynamic allocation.
Tenants may have SLA power guarantees, and server colors indicate tenant domains (see legend).
The dashed box highlights an example horizontal tenant SLA constraint coupling devices across the PDN, while the triangle highlights an example vertical hierarchical PDN constraint on a subtree.}
\label{fig:hierarchy}
\end{figure}
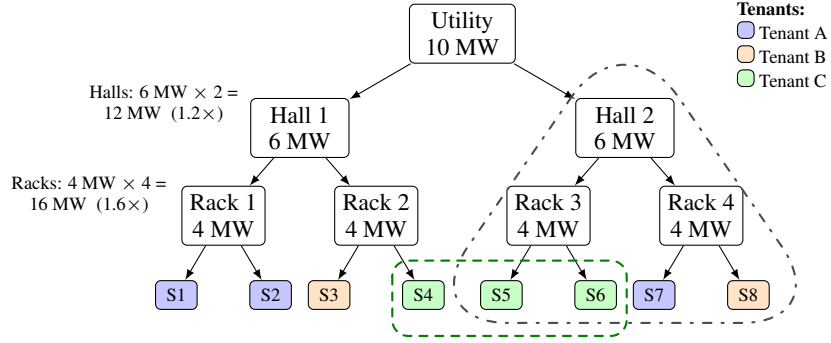

Figure~\ref{fig:hierarchy} illustrates a datacenter's power and tenant layout with multiple oversubscription levels and several tenant domains.
The problem is to design a policy that, given per-device predictions or requests, outputs device power allocations that satisfy hierarchical PDN constraints and tenant SLA constraints at every control step, respect job priorities and the active/idle distinction, and distribute shortage or surplus fairly.

\subsection*{Our Contribution}
To address these requirements in a single policy suitable for tight control loops, we make the following contributions:
\begin{itemize}
	\item \textbf{Problem formulation and requirements.}
	We formalize dynamic power allocation over a full tree-structured PDN topology with hierarchical capacity constraints and per-device power limits, horizontal tenant SLA constraints that couple devices across the tree, and operational requirements including job priorities and an active/idle distinction.
	
	\item \textbf{nvPAX: a hybrid QP/LP allocator with explicit fairness guarantees.}
	We propose nvPAX, a constrained-optimization policy realized as a hybrid sequence of convex programs: a priority-ordered Quadratic Program (QP) for power-request satisfaction followed by max--min Linear Programs (LP) for excess power redistribution among active and then idle devices.
	Our optimization formulation preserves deterministic feasibility at every control step while maximizing utilization and distributing shortage/surplus in a well-defined fair manner.
	Crucially, our hybrid optimization approach makes these large, constrained fairness problems solvable fast enough to run inside a tight control loop.
	
	\item \textbf{Saturation-aware surplus redistribution.}
	To avoid leaving usable power undistributed under tight PDN and tenant constraints, nvPAX iteratively detects saturated devices and re-solves reduced LPs, improving utilization without violating fairness or feasibility.
	
	\item \textbf{Support for heterogeneous device types.}
	We show how to incorporate per-device weights (i.e., supporting relative rather than absolute power allocations) so that fairness is meaningful across heterogeneous power ranges.
	
	\item \textbf{Baseline allocations and their limitations.}
	We describe a greedy baseline power allocation algorithm and show that, while it can match nvPAX on balanced hierarchies, it can be substantially inferior in non-uniform PDNs with internal bottlenecks; moreover, it cannot enforce horizontal tenant SLAs.
	
	\item \textbf{Large-scale evaluation and scalability.}
	Using GPU telemetry from a production datacenter with over $12{,}000$ H100 GPUs, we evaluate nvPAX against static and greedy baseline power allocations, and measure both allocation quality and runtime.
	On this trace, nvPAX attains a mean satisfaction ratio of 98.92\% while running within a tight control-loop budget (mean wall-clock time of 264.69~ms).
	We further benchmark scaling on synthetic hierarchies up to $10^5$ devices.
\end{itemize}

%% file: sections/related_work.tex
\section{Related Work}\label{related-work}

Datacenter power management aims to supply each power consumer with the power it needs while respecting physical limits, which include, for example, maximum power per rack or PDU so that circuit breakers are not triggered.
Device or server power capping is the practice of enforcing upper limits on power consumption to prevent overloading; when capping is triggered, device performance is typically reduced (e.g., by lowering processor frequency) so that consumption stays within the allowed budget.

Hierarchical power capping has been widely studied and deployed in production.
SHIP~\mbox{\cite{ship2009}} models the three-level power-distribution hierarchy (rack, PDU, datacenter) and uses a feedback control approach with formal stability guarantees.
Similarly, Dynamo~\mbox{\cite{wu2016dynamo}} coordinates power capping across Facebook's entire datacenter fleet by monitoring power at various hierarchy levels and reactively throttling lower-priority workloads during power emergencies.
Other works have explored managing distributed UPS energy for effective power capping~\mbox{\cite{kontorinis2015managing}} and agile, scalable datacenter power control~\mbox{\cite{miller2015scalable}}.
Further research addresses sustainable and carbon-aware computing~\mbox{\cite{bilge2023carbon,radovanovic2022carbon}}, emphasizing the need for robust power management.
However, these systems focus primarily on aggregate power limits rather than modeling per-device power requests.
Under power shortage, they rely on utilization-weighted proportional sharing (as in SHIP) or reactive heuristics (as in Dynamo), offering no formal guarantee of per-device fairness or strict adherence to tenant SLAs.

Priority-aware capping has been deployed at scale. Sakalkar et al.~\mbox{\cite{sakalkar20asplos}} describe Google's approach of pausing lower-priority workloads when power is short, so that production services keep their power. Kumbhare et al.~\mbox{\cite{kumbhare21atc}} use supervised machine learning (ML) predictions to protect critical virtual machines (VMs) on Azure.
Similarly, Thunderbolt~\mbox{\cite{li2020thunderbolt}} uses task-level QoS differentiation to throttle batch workloads while maintaining low latency for serving tasks.
More recently, power oversubscription has been explored specifically for modern AI workloads.
POLCA~\mbox{\cite{patel2023polca}} characterizes LLM power consumption to enable aggressive oversubscription in inference clouds, while TAPAS~\mbox{\cite{stojkovic2025tapas}} proposes thermal and power aware scheduling for LLM inference.
Other works investigate GPU sharing and fragmentation in deep learning clusters~\mbox{\cite{wu2023transparent}}.
Another solution~\mbox{\cite{dell2024fuzzy}} implements fuzzy-logic-based prioritized capping at the rack or server level, which remains heuristic and single-level.
None of these methods formulates allocation as a constrained optimization problem that simultaneously enforces hierarchical limits and tenant guarantees.

Optimization has been applied to datacenter power, but typically for different objectives.
Pelley et al.~\mbox{\cite{pelley2010power}} use optimization for dynamic power provisioning, but their focus is on dynamically altering the PDN topology (power routing) rather than fairly allocating a fixed hierarchical budget.
Linear programming has been used for energy-aware server provisioning~\mbox{\cite{guenter11managing}} (deciding how many servers to turn on or off to meet demand), but in a flat, single-cluster model and not for per-device power allocation.
Similarly, fast centralized schedulers like Firmament~\mbox{\cite{gog2016firmament}} and Apollo~\mbox{\cite{boutin2014apollo}} use network flow optimization or coordinated scheduling for task placement, but do not explicitly manage hierarchical power limits.
Finally, reinforcement learning and related methods~\mbox{\cite{parm23neurips,rldatacenter2025}} have been explored for power-aware resource management, but they generally lack deterministic feasibility guarantees at each control step.

Overall, each prior work only addresses a subset of the problem.
Modern datacenters are becoming extremely large and complex, with thousands or even tens of thousands of constraints that must be satisfied simultaneously~\mbox{\cite{barroso2013datacenter,jeon2019analysis}}.
Power oversubscription must therefore be managed in a way that satisfies all physical and service-level constraints at every moment while still providing high performance and efficient resource utilization.
Furthermore, tenant SLAs must be respected, and mechanisms for job prioritization and fair allocation are essential to avoid job bottlenecks, which are particularly detrimental to distributed deep neural network (DNN) and large language model (LLM) workloads~\mbox{\cite{gu2019tiresias,mahajan2020themis,xiao2018gandiva,narayanan2020heterogeneity,peng2018optimus,qiao2021pollux,zheng2023shockwave}}.

To the best of our knowledge, this paper is the first to jointly handle hierarchical PDN constraints, tenant SLA constraints, and per-device fairness in one allocation policy, realized as a sequence of convex optimization problems.
We support job priorities while distinguishing active and idle jobs, and every power allocation of our algorithm is guaranteed to satisfy all physical and SLA constraints.

%% file: sections/policy_requirements.tex
\section{Policy Requirements and Allocation Strategy}\label{sec:design-principles}

In this section, we outline the desired requirements from a power allocation policy.
Before doing so, we first define several key concepts:

\begin{itemize}
	\item \textbf{Active and idle devices:} An active device is one assigned to a running job; an idle device is not assigned to any running job.
	Device state can be classified using one of the following methods:
	\begin{enumerate}
		\item Job scheduler information: When accessible, the scheduler provides the most accurate indication of device activity, identifying whether a device is allocated to a running job or not.
		\item Power-based detection: In cloud environments, where the tenants' job-scheduler information is not available to the cloud service provider (CSP), device activity can be inferred from power consumption.
	\end{enumerate}
	\item \textbf{Power request:} A device’s power request may be derived from predicted power consumption (e.g., via a forecasting model such as PRS~\cite{sivan2024prs}), or from the job specification associated with the device, such as max-P, max-Q, or other power profiles~\cite{narayanaswamy2025datacenterenergyoptimizedpower}.
	The power request of an idle device equals the minimum allowed power limit.
	\item \textbf{Power allocation:} The power limit applied to a device.
\end{itemize}

The power allocation policy should satisfy the following requirements:

\begin{enumerate}
\item \textbf{Feasibility (physical and service-level constraints):}
The policy must satisfy all physical constraints (device power limits and hierarchical PDN capacities) and all service-level constraints (e.g., tenant SLAs) at every control step, and must never output an infeasible allocation.

\item \textbf{Closeness to requests:}
The policy should minimize the discrepancy between power requests and allocated power.
This ensures that allocations closely track actual needs while still respecting all physical and service-level constraints.

\item \textbf{Maximizing allocation of the available power budget:}
The policy should allocate as much of the available power budget as possible, subject to all constraints.
The objective should not ``reserve'' unused power budget.

\item \textbf{Idle vs.\ active prioritization:}
The policy must treat active and idle devices differently, ensuring that active devices get higher priority when distributing available power.
Idle devices should only receive increased power allocations (beyond required minimum) when surplus power remains, allowing active devices to receive sufficient power and headroom, while idle devices remain near their minimum permissible power.
At the same time, if surplus power remains after all active devices have been raised to their maximum allowed allocation, the allocation may also be increased for idle devices (since ``idle'' classification can be imperfect).

\item \textbf{Priority-aware allocation:}
The policy must respect per-device priorities.
Priority may reflect job importance, device type (e.g., networking devices such as switches may be assigned the highest priority to ensure fabric availability), or a combination of both.
It should first allocate power to satisfy the requests of the highest priority devices; only after all devices at this priority level are either fully satisfied or constrained should it proceed to the next priority level, and so on.
Once all per-device requests across all priority levels have been met, any remaining surplus is distributed (e.g., evenly) subject to constraints.
Priority information is optional; when unavailable, all devices are assigned equal priority and the allocator reduces to a single-level fair allocation.

\item \textbf{Fair handling of shortages and surplus power budget:}
In case of a shortage, the policy should allocate power such that lower-priority jobs are impacted first.
Within each priority level, deviations from requests should be distributed evenly across devices to avoid creating bottlenecks.
Once prioritized requests are satisfied, any remaining spare budget should be distributed fairly across active devices, and then to idle devices if surplus remains.

\item \textbf{Heterogeneous device types:}
When devices have significantly different power ranges, the objective should support per-device weights or normalization (e.g., by each device's maximum allowed power) so that ``fair'' deviation is meaningful across device types.

\item \textbf{Computational efficiency:}
The algorithm computing the power allocations should be sufficiently fast to support control loops that adjust power allocations every few seconds.

\end{enumerate}

To satisfy the requirements outlined above, we propose a three-phase allocation procedure:
\begin{enumerate}
	\item \textbf{Phase I: Priority-ordered request satisfaction.}
	Devices are grouped by priority and processed from the highest to the lowest priority level.
	At each priority level, the algorithm solves a constrained optimization problem that attempts to satisfy the power requests of devices
	at the current priority (that is, allocating power as close as possible to the power requested by the device), while fixing the allocations of higher-priority devices
	to their previously determined values and keeping lower-priority or idle devices at minimum (unless tenant lower-bound SLAs require higher allocations).
	
	\item \textbf{Phase II: Residual power redistribution among active devices.}
	After all priority levels have been processed, any remaining power budget is
	redistributed among active devices, increasing their allocations beyond the values assigned in
	Phase~I, subject to all constraints.
	
	\item \textbf{Phase III: Residual power redistribution among idle devices.}
	Finally, if residual power remains, it is allocated to idle devices while
	preserving feasibility with respect to all physical and tenant constraints.
\end{enumerate}

In the next section, we present \textbf{nvPAX}, a hybrid constrained-optimization instantiation of this three-phase procedure.
Phase~I is formulated as a convex QP for priority-ordered request satisfaction, while Phases~II and~III are LPs for max-min surplus redistribution, all subject to the physical and service-level constraints described above.
nvPAX is designed to be invoked by the datacenter power management system in a closed-loop control cycle (e.g., every 30~seconds); device failures and unexpected reductions in available power supply are therefore handled implicitly---at the next cycle, the management system passes updated device states and the current available capacity to nvPAX, which recomputes a feasible allocation from scratch.
Shortening the control interval reduces the window of exposure to any such disruption.

%% file: sections/algorithmic_framework.tex
\section{Algorithmic Framework}\label{sec:lp-minmax}

In this section, we derive our power allocation algorithm that satisfies all requirements stated in Section~\ref{sec:design-principles}.
We start by setting the required notation in Section~\ref{sec:notation}, and formulate the constraints in Section~\ref{sec:constraints}.
Then, in Section~\ref{sec:algorithm}, we present \textbf{nvPAX}, a hybrid QP/LP constrained-optimization approach: Phase~I uses a convex quadratic program (QP) for priority-ordered request satisfaction, while Phases~II and~III use linear programs (LPs) for max-min surplus redistribution.
This decomposition keeps all physical and tenant constraints linear while using the objective function best suited to each phase.

\subsection{Notation}\label{sec:notation}

We start by fixing the notation used by our algorithm.
We define notation for devices, their power limits and state, the power hierarchy and node capacities, and tenant domains and SLA budgets.
We denote by~$n$ the number of devices in our hierarchical PDN (such as GPUs, CPUs, etc.; in general, any power consumer whose power allocation can be controlled).

\begin{itemize}
	\item \textbf{Devices:} Devices are indexed by $i \in \{1, \ldots, n\}$.
	For each device $i$:
	\begin{itemize}
		\item $l_i, u_i$ denote its minimum and maximum allowed power limits.
		\item $a_i \in [l_i, u_i]$ denotes its allocated power (decision variable).
		\item $r_i \in [l_i, u_i]$ denotes its requested (or predicted) power, where for idle devices $r_i = l_i$.
		\item $p_i \in \{1, \ldots, P\}$ denotes its priority level, where $P$ is the highest priority level, and higher values indicate higher priority.
		In practice, $p_i$ reflects the relative importance of device~$i$ (see Section~\ref{sec:design-principles}); the priority of idle devices is irrelevant.
	\end{itemize}

	\item \textbf{Device states:} $\mathcal{R} \subseteq \{1, \ldots, n\}$ denotes the set of active devices, and $\mathcal{I} \subseteq \{1, \ldots, n\}$ denotes the set of idle devices, with $\mathcal{R} \cup \mathcal{I} = \{1, \ldots, n\}$ and $\mathcal{R} \cap \mathcal{I} = \emptyset$. If device~$i$ is idle, we set $r_{i} = l_{i}$.

	\item \textbf{Power hierarchy:} The physical power hierarchy forms a rooted tree.
	Nodes in the hierarchy are indexed by $j$.
	For each node $j$
	\begin{itemize}
		\item $C_j$ denotes the power capacity of node $j$,
		\item $\mathcal{D}_j \subseteq \{1, \ldots, n\}$ denotes the set of devices in the subtree rooted at node $j$.
	\end{itemize}

	\item \textbf{Tenants:} Tenants are indexed by $k$.
	Each tenant $k$ is associated with a set of devices $\mathcal{T}_k \subseteq \{1, \ldots, n\}$ and may impose SLA constraints on its allocations $\{a_i : i\in\mathcal{T}_k\}$.

\end{itemize}

\subsection{Constraints}
\label{sec:constraints}

Any power allocation $a_1, \ldots, a_n$ must satisfy the following constraints, independent of the objective function used to determine the allocation.

\begin{enumerate}
	\item \textbf{Device power limits:} Each device allocation must be within the device's physical lower and upper limits, that is
	\begin{equation}
		\label{eq:limits}
		l_i \le a_i \le u_i, \quad \forall i \in \{1, \ldots, n\}.
	\end{equation}

	\item \textbf{Hierarchical power constraints:} The total power allocated to all devices under a given node in the PDN must not exceed the node's power capacity, that is
	\begin{equation}
		\label{eq:hierarchy}
		\sum_{i \in \mathcal{D}_j} a_i \le C_j, \quad \forall j,
	\end{equation}
	where $j$ ranges over all nodes in the PDN.

	\item \textbf{Tenant SLA constraints:} We assume tenant/service-level constraints can be represented as linear inequalities.
	Each tenant constraint is of the form
	\begin{equation}
		\label{eq:sla}
		B_k^{\min} \;\le\; \sum_{i \in \mathcal{T}_k} a_i \;\le\; B_k^{\max}, \quad \forall k,
	\end{equation}
		where $B_k^{\min}\ge 0$ and $B_k^{\max}\ge 0$ are the guaranteed minimum and maximum power for tenant~$k$, respectively.
	Either bound may be absent (i.e., $B_k^{\min}=0$ or $B_k^{\max}=\infty$).
	More general linear SLA constraints (not shown) are also supported by our algorithm and can encode, e.g., per-device minimum guarantees or combinations of budgets across subsets of devices.
\end{enumerate}

\subsection{nvPAX}\label{sec:algorithm}

This section provides a detailed description of our power allocation algorithm.
It consists of three phases, where each phase is formulated as one or more convex optimization problems (either a QP or an LP) subject to the device, hierarchical, and tenant constraints of Section~\ref{sec:constraints}.
The result of the algorithm is a feasible allocation that respects priorities, maximizes utilization of the available power budget, and distributes surplus or shortage in a well-defined fair manner.

\subsubsection{\texorpdfstring{Phase I: Priority-Based Allocation}{Phase I: Priority-Based Quadratic Allocation}}

Phase~I allocates power to active devices as close as possible to per-device power requests while enforcing all constraints and respecting priority order.
Concretely, we process priorities from highest to lowest; within each priority level, we minimize the deviation $r_i-a_i$ for devices at that priority.

We initialize $a_i \gets l_i$ for all devices, then process priority levels one at a time from $P$ down to~1, completing each level before moving to the next; once allocations for a higher-priority level have been determined, they become hard constraints for all lower-priority levels.
At the start of level $p$, three disjoint device sets are initialized:
\begin{itemize}
	\item \textbf{Unsaturated set} $\mathcal{A} \gets \{i\in\mathcal{R} : p_i = p\}$: active devices at the current priority level whose allocations are being optimized in this iteration.
	\item \textbf{Fixed set} $\mathcal{F} \gets \{i\in\mathcal{R} : p_i > p\}$: higher-priority devices whose allocations were finalized in earlier iterations and are held fixed.
	\item \textbf{Free set} $\mathcal{L} \gets \{1,\ldots,n\}\setminus(\mathcal{A}\cup\mathcal{F})$: lower-priority active devices and idle devices, which are not yet being optimized but must remain flexible (within $[l_i,u_i]$) to satisfy SLA constraints.
\end{itemize}

For a fixed priority level, Phase~I solves the QP
\begin{equation}\label{eq:phase1_qp}
	\begin{aligned}
	\min_{\{a_i\}}\quad
	& \sum_{i\in\mathcal{A}} (a_i-r_i)^2
	  + \varepsilon \sum_{i\in\mathcal{L}} (a_i-l_i)^2 \\
	\text{s.t.}\quad
	& \text{constraints~\eqref{eq:limits}--\eqref{eq:sla}},\\
	& a_i = a_i^{\mathrm{fixed}}, \quad i\in\mathcal{F},\\
	& l_i \le a_i \le u_i, \quad i\in\mathcal{A}\cup\mathcal{L}.
	\end{aligned}
\end{equation}

Devices in $\mathcal{A}$ may receive allocations above their request $r_i$, and devices in $\mathcal{L}$ may receive allocations above their minimum $l_i$: in both cases, tenant SLA constraints of the form $\sum_{i\in\mathcal{T}_k} a_i \ge B_k^{\min}$ can force this, and fixing devices in $\mathcal{A}$ at $r_i$ or devices in $\mathcal{L}$ at $l_i$ could render power allocation infeasible.
The objective~\eqref{eq:phase1_qp} is designed to discourage unnecessary over-allocation: it pulls allocations of devices in $\mathcal{A}$ toward $r_i$ and allocations of devices in $\mathcal{L}$ toward $l_i$, deviating only as much as the constraints require. To that end, the first term in~\eqref{eq:phase1_qp} pulls current-priority devices toward their requests, while the second term keeps lower-priority and idle devices close to their minimum power unless tenant lower-bound SLAs force them upward. If no tenant lower-bound SLA is present, devices in $\mathcal{L}$ can be fixed at $l_i$ and the regularizer is unnecessary.

For heterogeneous devices (i.e., having substantially different power ranges), nvPAX also supports a normalized objective
\begin{equation*}\label{eq:phase1_qp_relative}
	\min_{\{a_i\}}\quad
	\sum_{i\in\mathcal{A}}
	\left((a_i-r_i) / u_i\right)^2
	+ \varepsilon \sum_{i\in\mathcal{L}}
	\left((a_i-l_i) / u_i \right)^2 ,
\end{equation*}
which normalizes each device's deviation by its maximum allowed power, making ``fair'' deviation comparable across devices with different power ranges.
More generally, $u_i$ may be replaced by any positive per-device scale.

The QP in~\eqref{eq:phase1_qp} is strictly convex over the free variables. When there are no lower bound SLA constraints, $\mathcal{L}$ is pinned and the free block is $\mathcal{A}$; with SLA lower bounds, $\varepsilon>0$ makes the Hessian positive on $\mathcal{A}\cup\mathcal{L}$. Variables in $\mathcal{F}$ are fixed by bounds, so any zero Hessian entries on fixed variables do not affect uniqueness of the solution. Thus, each priority level has a unique optimum and requires only a single QP solve.
The regularization parameter $\varepsilon$ is chosen small so that lower-priority and idle devices stay near their minimums unless feasibility requires otherwise; in practice the results are insensitive to the exact value, and we use $\varepsilon = 10^{-5}$, large enough to avoid numerical issues yet small enough not to distort the primary request-deviation objective.

Algorithm~\ref{alg:phase1} presents the procedure for Phase~I.

\begin{algorithm}[t]
\caption{Phase I: Priority-based request satisfaction}
\label{alg:phase1}
\begin{algorithmic}[1]
\REQUIRE Devices $\{(r_i, l_i, u_i, p_i)\}_{i=1}^{n}$; PDN $\{(C_j, \mathcal{D}_j)\}$; tenant SLAs $\{(B_k^1, B_k^2, \ldots, \mathcal{T}_k)\}$
\ENSURE Allocation $\{a_i\}$ satisfying constraints~\eqref{eq:limits}--\eqref{eq:sla}
\STATE Initialize $a_i = l_i$ for all $i$
\FOR{each positive priority level $p$ from highest to lowest}
    \STATE $\mathcal{A} \gets \{i\in\mathcal{R} : p_i = p\}$, \quad $\mathcal{F} \gets \{i\in\mathcal{R} : p_i > p\}$, \quad $\mathcal{L} \gets \{1,\ldots,n\}\setminus(\mathcal{A}\cup\mathcal{F})$
    \STATE $a^* \gets$ solve~\eqref{eq:phase1_qp} with sets $\mathcal{A}$, $\mathcal{F}$, $\mathcal{L}$
    \STATE $a_{i} \gets a^*_{i}$ for all $i$
\ENDFOR
    \RETURN $\{a_{i}\}$
\end{algorithmic}
\end{algorithm}

\subsubsection{Phase II: Max--Min Allocation for Active Devices}

After Phase I, remaining power is distributed to active devices using a max-min fairness objective.
For Phase~II, devices are partitioned into three sets~$\mathcal{A}$, $\mathcal{F}$, and $\mathcal{L}$, initialized by
\[
\mathcal{A} \gets \mathcal{R},\qquad
\mathcal{F} \gets \emptyset, \qquad \mathcal{L} \gets \mathcal{I}.
\]
Here $\mathcal{A}$ contains the active devices to be optimized and the free set $\mathcal{L}$ contains the idle devices.

In Phase II, we maximize the minimum additional power allocated across all active devices in $\mathcal{A}$ by solving
\begin{equation*}
	\begin{aligned}
		\max_{\{a_i\}} \min_{i \in \mathcal{A}}\quad & a_i - a_i^{(1)},
	\end{aligned}
\end{equation*}
which, via the standard epigraph transformation and a regularization term, is written as the LP
\begin{equation}\label{eq:phase2LP}
\begin{aligned}
	\max_{\{a_i\},\, t}\quad & t \;+\; \varepsilon\!\sum_{i\in\mathcal{A}} a_i \;-\; \varepsilon\!\sum_{i\in\mathcal{L}} a_i \\
	\text{s.t.}\quad & t \ge 0, \\
				  & a_i - a_i^{(1)} \ge t, \quad \forall i \in \mathcal{A},\\
				  & a_i = a_i^{\text{fixed}}, \quad \forall i \in \mathcal{F},
\end{aligned}
\end{equation}
subject to constraints~\eqref{eq:limits}--\eqref{eq:sla}, where $a_i^{(1)}$ denotes the Phase I allocation. For heterogeneous devices, we can use normalized improvement by replacing $a_i-a_i^{(1)}\ge t$ with $(a_i-a_i^{(1)})/u_i\ge t$.

The $+\varepsilon$ coefficient on $\mathcal{A}$ selects, among all optimal solutions, one that maximizes total active-device allocation, promoting higher utilization.
The $-\varepsilon$ coefficient on $\mathcal{L}$ pushes free idle devices toward~$l_i$; the LP raises them only when tenant SLA constraints require it.
As in Phase~I, $\varepsilon$ is chosen small so that the max-min surplus objective dominates.

After solving~\eqref{eq:phase2LP} and increasing the allocation of each active device by~$t$, some devices may become saturated, namely, their allocation cannot be increased without violating some constraint (either its own upper bound, a tight ancestor-node capacity, the root capacity, or a tenant upper budget containing that device). We thus solve a sequence of LPs, where at each iteration we fix the allocation of saturated devices and re-solve for the remaining devices. 
Algorithm~\ref{alg:lp-phase2} summarizes the procedure of Phase II.

\begin{algorithm}[t]
\caption{Phase II: Surplus allocation for active devices}
\label{alg:lp-phase2}
\begin{algorithmic}[1]
\REQUIRE Phase~I allocation $\{a_i^{(1)}\}$; states $\mathcal{R}, \mathcal{I}$; PDN $\{(C_j, \mathcal{D}_j)\}$; tenant SLAs $\{(B_k^1, B_k^2, \ldots, \mathcal{T}_k)\}$
\ENSURE Updated allocation $\{a_i\}$ satisfying constraints~\eqref{eq:limits}--\eqref{eq:sla}
\STATE $\mathcal{A} \gets \mathcal{R}$, \quad $\mathcal{F} \gets \emptyset$, \quad $\mathcal{L} \gets \mathcal{I}$
\WHILE{$\mathcal{A} \neq \emptyset$}
    \STATE $(a^*, t^{*}) \gets $ solve~\eqref{eq:phase2LP} with sets $\mathcal{A}$, $\mathcal{F}$, $\mathcal{L}$
    \STATE $a_{i} \gets a^*_i$ for all $i$
	\STATE Identify saturated devices: $\mathcal{N} \gets \{i \in \mathcal{A} : i\text{ has no positive slack to receive more surplus}\}$
    \STATE $\mathcal{A} \gets \mathcal{A} \setminus \mathcal{N}$
    \STATE $\mathcal{F} \gets \mathcal{F} \cup \mathcal{N}$ \COMMENT{$\mathcal{L}$ unchanged: $\mathcal{A}\cup\mathcal{F}$ is invariant}
\ENDWHILE
\RETURN $\{a_{i}\}$
\end{algorithmic}
\end{algorithm}

\subsubsection{Phase III: Progressive Allocation for Idle Devices}

Phase III distributes any remaining surplus power to idle devices, using the same max-min procedure as Phase~II with $\mathcal{A}\gets\mathcal{I}$ and $\mathcal{F}\gets\mathcal{R}$.
Active devices are pinned at their Phase~II allocations via $a_i = a_i^{(2)}$ for all $i\in\mathcal{F}$.
There is no free set~$\mathcal{L}$ in this phase: all constraints, including tenant minimum SLAs, were already satisfied at the end of Phase~II, and since Phase~III only increases idle allocations, all tenant lower bounds are maintained or improved.

The LP of Phase III maximizes the minimum power increase of idle devices, with a regularization term that promotes higher total allocation:
\begin{equation}\label{eq:phase3LP}
\begin{aligned}
	\max_{\{a_i\},\, t}\quad & t \;+\; \varepsilon\!\sum_{i\in\mathcal{A}} a_i \\
	\text{s.t.}\quad & t \ge 0, \\
				  & a_i - a_i^{(2)} \ge t, \quad \forall i \in \mathcal{A}, \\
				  & a_i = a_i^{(2)}, \quad \forall i \in \mathcal{F},
\end{aligned}
\end{equation}
subject to constraints~\eqref{eq:limits}--\eqref{eq:sla}, where $a_i^{(2)}$ denotes the Phase~II allocation.

The $+\varepsilon$ coefficient selects, among all optimal solutions, one that maximizes total idle allocation.
As in the preceding phases, we solve a sequence of LPs with iterative saturation detection.
The procedure for Phase~III is the same as Algorithm~\ref{alg:lp-phase2}, but starting from the Phase~II allocations $\{a_i^{(2)}\}$, while setting $\mathcal{A}\gets\mathcal{I}$, $\mathcal{F}\gets\mathcal{R}$, $\mathcal{L}\gets\emptyset$, and using LP~\eqref{eq:phase3LP} in place of LP~\eqref{eq:phase2LP}.

Algorithm~\ref{alg:nvpax} summarizes the complete nvPAX procedure by composing the three phases described above.

\begin{algorithm}[t]
\caption{nvPAX: Three-phase power allocation}
\label{alg:nvpax}
\begin{algorithmic}[1]
\REQUIRE Devices $\{(r_i, l_i, u_i, p_i)\}_{i=1}^{n}$; states $\mathcal{R}, \mathcal{I}$; PDN $\{(C_j, \mathcal{D}_j)\}$; SLAs $\{(B_k^1, B_k^2, \ldots, \mathcal{T}_k)\}$
\ENSURE Allocation $\{a_i\}$ satisfying constraints~\eqref{eq:limits}--\eqref{eq:sla}
\STATE \textbf{Phase I (priority-ordered request satisfaction):} $\{a_i^{(1)}\} \gets$ Algorithm~\ref{alg:phase1} on $(r_i,l_i,u_i,p_i)$
\STATE \textbf{Phase II (redistribute among active devices):} $\{a_i^{(2)}\} \gets$ Algorithm~\ref{alg:lp-phase2} given $\{a_i^{(1)}\}$, with $\mathcal{A}\gets\mathcal{R}, \mathcal{F}\gets\emptyset, \mathcal{L}\gets\mathcal{I}$
\STATE \textbf{Phase III (redistribute among idle devices):} $\{a_i^{(3)}\} \gets$ Algorithm~\ref{alg:lp-phase2} given $\{a_i^{(2)}\}$, with $\mathcal{A}\gets\mathcal{I}, \mathcal{F}\gets\mathcal{R}, \mathcal{L}\gets\emptyset$
\RETURN $\{a_i^{(3)}\}$
\end{algorithmic}
\end{algorithm}

%

%% file: sections/experiments.tex
\section{Experiments}\label{sec:numerical-results}

We evaluate nvPAX on a large-scale simulation constructed from GPU power telemetry from a production datacenter.
We compare its power allocations against static equal-share and greedy proportional baselines explained below.

All experiments were run on an Apple M4 Pro with 14 cores and 48~GiB RAM. The software stack was Python~3.13.9, \texttt{SciPy}~1.16.3, HiGHS LP solver implemented by \texttt{highspy} Python bindings (version~1.12.0), and Clarabel (from \texttt{clarabel} version 0.11.1) for the Phase~I QP.
This reflects a deliberate design choice: nvPAX runs entirely on CPU and requires no GPU on the management node, keeping the power-management control plane independent of the accelerator hardware it manages.

\subsection{Data and Hierarchy}
The dataset comprises H100 GPU power metrics sampled every 30~seconds over a three-day period (8{,}523 timestamps).
The datacenter consists of 4 halls, with more than 12,000 GPUs.
The power hierarchy is: datacenter $\to$ halls $\to$ racks $\to$ servers $\to$ devices.
Each device is one GPU with $l_i = 200$~W and $u_i = 700$~W, consistent with H100 specifications.
Capacities are computed bottom-up.
At each server, capacity equals number of GPUs per server multiplied by 700~W (no oversubscription at server level).
At each rack, capacity equals the sum of its child server capacities multiplied by an oversubscription factor; at each hall, the sum of its rack capacities multiplied by the oversubscription factor; and at the root (datacenter), the sum of hall capacities multiplied by the oversubscription factor.
Thus at every level the parent's capacity is a fraction of the aggregate child capacity, producing hierarchical oversubscription. A single oversubscription factor (0.85 in the reported runs) is applied at each level, i.e., the parent's capacity at each level equals the sum of its children's capacities multiplied by this factor.
Hence, the parent can supply only that fraction of the aggregate child capacity.
The resulting ratio of total device maximum power to root capacity is approximately 1.63 in our hierarchy, so the network is undersupplied relative to all devices at maximum power, motivating dynamic allocation.

\subsection{Request Generation}
For each timestamp, each device's power request $r_i$ is the measured GPU power from the available telemetry for that GPU.
In this evaluation, we use the actual measured power as the request.
Before optimization, requests are clipped to the feasible device interval $[l_i,u_i]$; thus devices whose measured power is below $l_i$ enter the optimizer with request $l_i$, and similarly for $u_{i}$.
We assume perfect knowledge of device requests to evaluate the policy's ability to satisfy them under constraints.
A~device is classified as idle if its power request is strictly below 150~W; otherwise it is active.
This classification drives the Phase~II/III distinction (redistribution to active devices, then to idle devices).
In this experiment, all devices have priority 1. An experiment with multiple priority levels is described below.

\subsection{Baseline Allocations}
We compare \textbf{nvPAX} of Algorithm~\ref{alg:nvpax} against the following baseline allocations:
\begin{itemize}
	\item \textbf{Static allocation:} Each device receives an equal share of the root budget ($C_{\text{root}}/n$) with no redistribution of unused power.
	\item \textbf{Greedy proportional allocation:} A hierarchical heuristic that mimics industry-standard proportional sharing (similar to SHIP~\mbox{\cite{ship2009}}).
	Starting from the root, if a node is oversubscribed, it distributes its power to its children proportional to their aggregate requests, repeating recursively down to the devices.
	While this approach is fast and intuitive, it does not globally redistribute leftover power nor explicitly handles horizontal tenant SLA constraints.
\end{itemize}
Tenant SLAs introduce horizontal constraints that couple devices across the physical hierarchy (e.g., $\sum_{i\in\mathcal{T}_k} a_i \ge B_k$).
Since the static and greedy baselines do not enforce such constraints, they can violate tenant SLAs even when a feasible allocation exists.
Therefore, in the evaluation on real GPU telemetry we do not impose tenant SLA constraints, and compare policies only under device limits and hierarchical PDN constraints.
An experiment with tenant-SLA constraints and multiple job priorities is given in Appendix~\ref{app:sla-experiment}.

\subsection{Evaluation Metrics}

We compute the following quantities at each timestamp $t$.
Let $n$ denote the total number of devices, and let
$U^{\mathrm{nvPAX}} = \sum_{i=1}^{n} \min(r_i, a_i^{\mathrm{nvPAX}})$
be the total power allocated by nvPAX, capped by request, at timestamp $t$.
We define \(U^{\mathrm{Static}}\) and \(U^{\mathrm{Greedy}}\) analogously by replacing \(a_i^{\mathrm{nvPAX}}\) with \(a_i^{\mathrm{Static}}\) and \(a_i^{\mathrm{Greedy}}\), respectively. We use the following metrics:

\begin{enumerate}

\item \textbf{Relative utilization improvement (nvPAX vs.\ baselines, \%):}
The per-timestamp relative improvement is given by
\[
\Delta U^{\mathrm{Static}} = \frac{U^{\mathrm{nvPAX}} - U^{\mathrm{Static}}}{U^{\mathrm{Static}}} \cdot 100,
\qquad
\Delta U^{\mathrm{Greedy}} = \frac{U^{\mathrm{nvPAX}} - U^{\mathrm{Greedy}}}{U^{\mathrm{Greedy}}} \cdot 100.
\]
It measures how much nvPAX improves over each baseline as a percentage of that baseline's utilization.

\item \textbf{Satisfaction ratio:}
\[
S^{\mathrm{nvPAX}} = \frac{U^{\mathrm{nvPAX}}}{\sum_{i=1}^{n} r_i},
\]
defined when $\sum_{i=1}^{n} r_i > 0$.
This is the fraction of aggregate demand that is actually met; $S = 1$ means every device receives at least its requested power, while $S < 1$ quantifies the overall shortfall.
We define \(S^{\mathrm{Static}}\) and \(S^{\mathrm{Greedy}}\) analogously by replacing \(U^{\mathrm{nvPAX}}\) with \(U^{\mathrm{Static}}\) and \(U^{\mathrm{Greedy}}\), respectively.

\end{enumerate}

\subsection{Results}

\begin{figure*}
	\centering
	\begin{subfigure}[b]{0.48\textwidth}
		\centering
		\includegraphics[width=\textwidth]{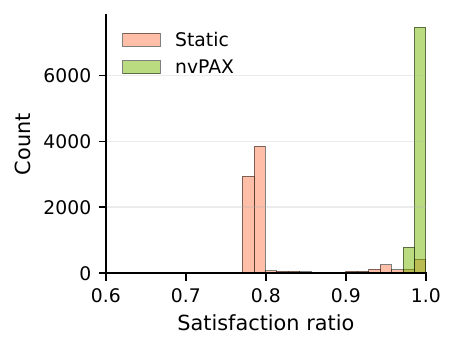}
		\caption{Satisfaction ratio of nvPAX ($S^{\mathrm{nvPAX}}$) and Static ($S^{\mathrm{Static}}$). Higher is better.}
		\label{fig:numerical-results-sat}
	\end{subfigure}
	\hfill
	\begin{subfigure}[b]{0.48\textwidth}
		\centering
		\includegraphics[width=\textwidth]{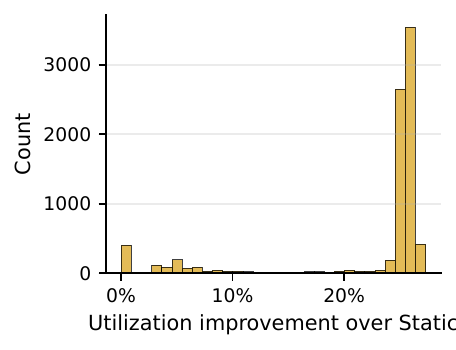}
		\caption{Relative utilization improvement of nvPAX over Static ($\Delta U^{\mathrm{Static}}$).}
		\label{fig:numerical-results-util-rel-static}
	\end{subfigure}
	\caption{nvPAX vs.\ Static equal-share allocation. Left: satisfaction ratio. Right: relative utilization improvement over Static. Overall, nvPAX achieves substantially higher satisfaction than Static and delivers higher useful utilization, indicating that it allocates more of the available power budget to meet demand under the constraints.}
	\label{fig:numerical-results-main}
\end{figure*}

We ran the simulation for a duration corresponding to three consecutive days of telemetry (8{,}523 timestamps at 30-second spacing).
Figure~\ref{fig:numerical-results-main} summarizes nvPAX vs.\ Static: satisfaction ratio (left) and relative utilization improvement over Static (right).
Greedy proportional allocation is not shown since on this trace its aggregate satisfaction is almost identical to nvPAX (mean $S^{\mathrm{Greedy}} = 98.92\%$ vs.\ $S^{\mathrm{nvPAX}} = 98.92\%$), so the nvPAX--Greedy utilization gap is negligible; both policies substantially outperform Static under hierarchical oversubscription. For satisfaction ratio, nvPAX achieved mean 98.92\% (std 0.48\%, min 96.49\%, max 100.00\%); Static 81.30\% (std 6.37\%, min 77.07\%, max 100.00\%); Greedy 98.92\% (std 0.48\%, min 96.49\%, max 100.00\%).
The mean improvement of nvPAX over Static was $17.62$ percentage points (std 5.93\%); nvPAX was at least as good as Static on every timestamp.

Despite the similar performance of nvPAX and the Greedy algorithm in this simulation, they are not competitive in general.
Since the Greedy algorithm makes local proportional decisions based on aggregate subtree demand, it may allocate budget to subtrees whose internal PDN bottlenecks prevent that budget from being delivered to devices.
In such non-uniform hierarchies, nvPAX’s global constrained-optimization formulation reallocates power away from bottlenecked subtrees toward feasible headroom elsewhere, yielding substantially higher utilization and satisfaction. Appendix~\ref{app:greedy} constructs a concrete example where the Greedy algorithm is considerably inferior to nvPAX and characterizes when this behavior arises. Finally, nvPAX is strictly more expressive than Static allocation and the Greedy algorithm in that it enforces horizontal tenant SLA constraints that couple devices across the PDN. This capability is absent from these baseline algorithms.

Runtime of nvPAX in this simulation (single \texttt{optimize()} call per timestamp): mean 264.69~ms with std~162.72~ms.

\subsection{Runtime and Computational Complexity}

\begin{figure}[t]
	\centering
	\includegraphics[width=0.48\textwidth]{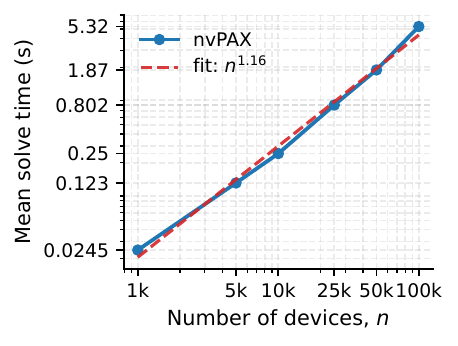}
	\caption{Empirical scaling of nvPAX's optimization time on synthetic random hierarchies. }
	\label{fig:benchmark-timing}
\end{figure}

To estimate the runtime of nvPAX and its scaling with problem size, we benchmark its performance on synthetic randomly generated hierarchies with device counts $n \in \{10^3, 5{\cdot}10^3, 10^4, 2.5{\cdot}10^4, 5{\cdot}10^4, 10^5\}$, with five independent runs per size.
We measure the wall-clock time of a single \texttt{optimize()} call (including all phases).
Figure~\ref{fig:benchmark-timing} shows the results.

\paragraph{Scalability and Runtime.}
We note that the presented benchmark computes per-device power allocations.
In newer and emerging platforms (e.g., GB200-class servers and successors), power caps may be applied per server (typically covering 4-8 GPUs and several CPUs) rather than per device; this reduces the number of decision variables and constraints and can therefore yield substantially faster runtimes.
Extrapolating from Figure~\ref{fig:benchmark-timing}, we expect that solving for \(\sim\)10{,}000 servers can be achieved in under 0.25 seconds.
Additional speedups are possible via early stopping of surplus distribution in Phases~II and~III, warm-starting across control steps, and decomposing the hierarchy into independent subproblems when constraints decouple.
Finally, faster runtimes may be achievable with commercial QP/LP solvers (e.g., Gurobi~\cite{gurobi}) and stronger CPUs.

\paragraph{Computational Complexity.}
Phase~I solves one convex QP per priority level, while Phases~II and~III solve a sequence of LPs with iterative saturation detection. 
Each QP/LP has $O(n)$ allocation variables (plus an auxiliary scalar $t$ in Phases~II and~III) and $O(n + N_{\mathrm{nodes}} + N_{\mathrm{SLA}})$ linear constraints for device bounds and physical and tenant constraints, with sparse connectivity.
Thus a single solve is polynomial in the problem dimensions; practical cost is dominated by the QP/LP solver calls.
Figure~\ref{fig:benchmark-timing} shows that on the synthetic hierarchies, observed mean runtime scales approximately as $n^{1.16}$ over $10^3$--$10^5$ devices.

%% file: sections/conclusion.tex
\section{Conclusion}\label{sec:conclusion}

This paper presented a hybrid QP/LP constrained-optimization algorithm for dynamic power allocation in hierarchical, multi-tenant datacenters.
The algorithm extends existing power capping frameworks by supporting a full tree-structured PDN and tenant SLA constraints (e.g., minimum and maximum power guarantees per tenant), while satisfying device limits and delivering priority-aware, fair allocation.
Empirical results on a large-scale simulation with thousands of GPUs and real telemetry, comparing nvPAX to Static equal-share and Greedy proportional baselines, demonstrate substantial gains in utilization and request satisfaction, together with empirical scaling behavior on synthetic hierarchies up to $10^5$ devices.

For future work, we plan to integrate nvPAX into production datacenter power-management systems such as PRS~\cite{sivan2024prs} and DPS~\cite{dps}.
Beyond deployment, an important direction is to characterize and improve the closed-loop behavior of nvPAX under noisy telemetry and rapidly changing demand, including adding explicit stability/smoothness mechanisms that limit unnecessary allocation oscillations across control steps.
A complementary direction is to develop an anytime, deadline-aware variant of nvPAX that returns progressively better allocations as time permits, while providing principled fallback when strict per-step latency constraints truncate later refinement phases.
Finally, we plan to exploit the spatial locality of demand changes in large hierarchies, aiming to confine allocation updates to affected subtrees and reduce disruption to unrelated devices.

%% file: sections/appendix.tex
\appendix

\newpage
\section{Greedy Proportional Allocation vs.\ nvPAX}
\label{app:greedy}

This appendix describes the Greedy proportional allocation baseline used in our experiments, and explains its limitations as a solution to the power allocation problems.

Greedy proportional allocation is a fast, top-down heuristic: at each node in the hierarchy, the available budget is split among children in proportion to their aggregate demand, recursing until every device has an allocation.
Because each split uses only local subtree information, the algorithm runs in a single pass through the tree and is easy to implement.
However, it has fundamental limitations compared to a global optimizer such as nvPAX:
\begin{itemize}
	\item \textbf{No cross-subtree coordination.}
	Greedy allocation considers each parent-to-child split independently, without considering capacity bottlenecks deeper or in other branches.
	When demand and capacity are distributed non-uniformly across the tree, a parent may allocate too much budget to a subtree that cannot use it (due to tight internal nodes) and too little to subtrees that could.
	Global allocation schemes such as nvPAX avoid this by solving a single optimization problem over all devices and constraints simultaneously.
	\item \textbf{No tenant SLA support.}
	The Greedy allocation follows the tree structure only and cannot encode horizontal constraints such as per-tenant minimum power guarantees.
	\item \textbf{No fairness objective.}
	The Greedy allocation distributes budget proportionally to demand without an explicit fairness criterion.

\end{itemize}
In balanced hierarchies where every node has ample headroom relative to its children's demands, the Greedy allocation and nvPAX produce similar allocations (as observed in Section~\ref{sec:numerical-results}).
The gap between the two methods becomes significant when the hierarchy contains non-uniform capacities or demand distributions, as shown in Section~\ref{sec:greedy-nonuniform} below.

\subsection{Greedy Proportional Algorithm}

In this section, we describe in detail the Greedy allocation algorithm. 
Each device~$i$ has bounds $[l_i,u_i]$; requests are clipped to $d_i \in [l_i,u_i]$.
Write $e_i = d_i - l_i \ge 0$ for extra demand above the minimum, and initialize $a_i \leftarrow l_i$.
For each node~$v$ in the hierarchy, let $L_v$ be the sum of $l_i$ over devices in the subtree rooted at~$v$, and $E_v$ the sum of $e_i$ over the same subtree.
Let $C_v$ be $v$'s power capacity.
Define the extra capacity $X_v = \max\{0, C_v - L_v\}$ and the feasible extra weight $W_v = \min(E_v, X_v)$, which is the maximum additional power above the minimal allocation  that the subtree can carry without violating $v$'s cap.
The values $(L_v,E_v,W_v)$ are computed bottom-up.

The root receives extra budget $W_{\mathrm{root}}$ (equivalently, one distributes total power $\min(C_{\mathrm{root}}, L_{\mathrm{root}} + W_{\mathrm{root}})$).
At each node~$v$, any positive extra budget is split among the immediate children (child nodes and devices attached to~$v$) in proportion to their respective weights: $W_c$ for a child node~$c$, and $e_i$ for a device~$i$ attached directly to~$v$.
Each child node recursively receives its share as its extra budget; each device receives its share added to~$a_i$.
This top-down pass respects all node capacities by construction of~$W_v$, but does not encode tenant SLA constraints.
The Greedy allocation algorithm is summarized in Algorithm~\ref{alg:greedy-prop}.

\begin{algorithm}[t]
\caption{Greedy proportional allocation.}
\label{alg:greedy-prop}
\begin{algorithmic}[1]
\REQUIRE Devices $(r_i, l_i, u_i)_{i=1}^{n}$; PDN hierarchy with node capacities $C_v$
\ENSURE Allocation $\{a_i\}_{i=1}^{n}$
\STATE \textbf{Initialization:}
\FOR{each device $i$}
	\STATE $d_i \gets \min(\max(r_i, l_i), u_i)$ \COMMENT{clip request to $[l_i, u_i]$}
	\STATE $e_i \gets d_i - l_i$ \COMMENT{extra demand above minimum}
	\STATE $a_i \gets l_i$ \COMMENT{allocate minimum}
\ENDFOR
\STATE
\STATE \textbf{Bottom-up aggregation:}
\FOR{each node $v$ in post-order (leaves first)}
	\STATE $L_v \gets \sum_{i \in \text{subtree}(v)} l_i$ \COMMENT{sum of device minimums}
	\STATE $E_v \gets \sum_{i \in \text{subtree}(v)} e_i$ \COMMENT{sum of extra demands}
	\STATE $X_v \gets \max\{0,\; C_v - L_v\}$ \COMMENT{extra capacity above minimums}
	\STATE $W_v \gets \min(E_v,\; X_v)$ \COMMENT{feasible extra weight}
\ENDFOR
\STATE
\STATE \textbf{Top-down distribution:}
\STATE Call \textsc{Distribute}$(\text{root},\; W_{\text{root}})$
\end{algorithmic}
\end{algorithm}

\begin{algorithm}[t]
\caption{\textsc{Distribute}$(v, b)$: recursive top-down budget distribution for node $v$ with extra budget $b$.}
\label{alg:greedy-distribute}
\begin{algorithmic}[1]
\REQUIRE Node $v$; extra budget $b \ge 0$
\IF{$b \le 0$}
	\RETURN
\ENDIF
\STATE $W_{\mathrm{tot}} \gets \sum_{c \in \mathrm{children}(v)} W_c + \sum_{i \in \mathrm{devices}(v)} e_i$
\IF{$W_{\mathrm{tot}} = 0$}
	\RETURN
\ENDIF
\FOR{each child node $c$ of $v$}
	\STATE $b_c \gets \min\!\bigl(b \cdot W_c / W_{\mathrm{tot}},\; W_c\bigr)$ \COMMENT{proportional share, capped}
	\STATE \textsc{Distribute}$(c, b_c)$ \COMMENT{recurse into subtree}
	\STATE $b \gets b - b_c$;\; $W_{\mathrm{tot}} \gets W_{\mathrm{tot}} - W_c$
\ENDFOR
\FOR{each device $i$ attached directly to $v$}
	\STATE $s_i \gets \min\!\bigl(b \cdot e_i / W_{\mathrm{tot}},\; e_i\bigr)$ \COMMENT{proportional share, capped}
	\STATE $a_i \gets a_i + s_i$
	\STATE $b \gets b - s_i$;\; $W_{\mathrm{tot}} \gets W_{\mathrm{tot}} - e_i$
\ENDFOR
\end{algorithmic}
\end{algorithm}

\subsection{Non-Uniform Hierarchy}
\label{sec:greedy-nonuniform}

We now construct a small example in which the Greedy algorithm's local decisions lead to a substantially worse allocation compared to nvPAX's global optimum.
The failure arises whenever demand and internal node capacities are distributed non-uniformly across the tree, so that a subtree with high aggregate demand contains a tight internal bottleneck that prevents most of that demand from being satisfied.

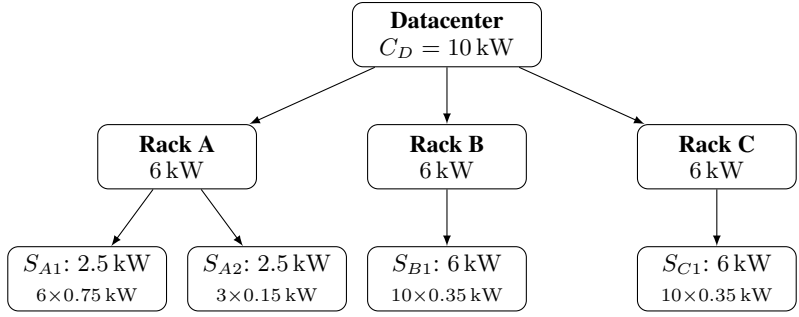
\begin{figure}[t]
	\centering
	\begin{tikzpicture}[
		scale=0.85,
		x=1cm, y=1cm,
		box/.style={
			draw,
			rounded corners,
			align=center,
			inner sep=2pt,
			font=\small,
			minimum width=2.1cm,
			minimum height=0.85cm,
			anchor=center,
		},
		edge/.style={draw, -latex},
		]
		\node[box, minimum width=2.5cm] (D) at (0,0)
		{\textbf{Datacenter}\\$C_D=10\,\mathrm{kW}$};

		\node[box] (RA) at (-4.2,-1.9) {\textbf{Rack A}\\$6\,\mathrm{kW}$};
		\node[box] (RB) at (0,-1.9)   {\textbf{Rack B}\\$6\,\mathrm{kW}$};
		\node[box] (RC) at (4.2,-1.9) {\textbf{Rack C}\\$6\,\mathrm{kW}$};

		\draw[edge] (D) -- (RA);
		\draw[edge] (D) -- (RB);
		\draw[edge] (D) -- (RC);

		\node[box] (SA1) at (-5.6,-3.8)
		{$S_{A1}$: $2.5\,\mathrm{kW}$\\[1pt]
			\scriptsize $6{\times}0.75\,\mathrm{kW}$};
		\node[box] (SA2) at (-2.8,-3.8)
		{$S_{A2}$: $2.5\,\mathrm{kW}$\\[1pt]
			\scriptsize $3{\times}0.15\,\mathrm{kW}$};

		\draw[edge] (RA) -- (SA1);
		\draw[edge] (RA) -- (SA2);

		\node[box] (SB) at (0,-3.8)
		{$S_{B1}$: $6\,\mathrm{kW}$\\[1pt]
			\scriptsize $10{\times}0.35\,\mathrm{kW}$};
		\node[box] (SC) at (4.2,-3.8)
		{$S_{C1}$: $6\,\mathrm{kW}$\\[1pt]
			\scriptsize $10{\times}0.35\,\mathrm{kW}$};

		\draw[edge] (RB) -- (SB);
		\draw[edge] (RC) -- (SC);
	\end{tikzpicture}

	\caption{Non-uniform hierarchy example.
	Rack~A has a tight internal server $S_{A1}$ ($2.5\,\mathrm{kW}$ capacity vs.\ $4.5\,\mathrm{kW}$ demand), while racks B and C have ample server capacity.
	Greedy proportional allocation over-allocates to rack~A at the datacenter level because it sees high aggregate demand, but much of that budget is wasted at the $S_{A1}$ bottleneck.}
	\label{fig:asymmetric-greedy}
\end{figure}

Consider the three-rack hierarchy in Figure~\ref{fig:asymmetric-greedy}: the datacenter cap is
$10\,\mathrm{kW}$ and total requested power is $11.95\,\mathrm{kW}$.
Rack~A contains a tight internal server $S_{A1}$ (capacity $2.5\,\mathrm{kW}$) feeding six devices that each
request $0.75\,\mathrm{kW}$ (total demand $4.5\,\mathrm{kW}$, but at most $2.5\,\mathrm{kW}$ can be delivered through $S_{A1}$).
A second server $S_{A2}$ under rack~A carries three devices at $0.15\,\mathrm{kW}$ each.
Racks B and C are symmetric: each has one $6\,\mathrm{kW}$ server and ten devices at $0.35\,\mathrm{kW}$.
All $29$ devices are marked \emph{active} with priority~1
so that Greedy does not use priority tiers -- any gap is purely from the
hierarchical proportional rule vs.\ the global optimum.

At the datacenter level, the Greedy algorithm splits the $10\,\mathrm{kW}$ budget among the three racks proportionally to their feasible extra weights $W_v$.
Rack~A reports high demand through its computed~$W_v$, so it receives a large share.
However, internally, most of that budget cannot pass through the tight server~$S_{A1}$: the six high-demand devices under $S_{A1}$ can receive at most $2.5\,\mathrm{kW}$ regardless of how much rack~A is given.
The surplus allocated to rack~A is therefore wasted, while racks B and C, whose servers have ample capacity, receive less than they could use.

Specifically, we evaluated satisfaction ratio $S = \sum_i \min(r_i,a_i) / \sum_i r_i$ with identical requests for both nvPAX and the Greedy algorithm.
While nvPAX achieves $S = 83.26\%$, the Greedy algorithm achieves $S = 73.94\%$, a gap of $+9.32$ percentage points.
nvPAX recognizes that rack~A's demand is largely undeliverable and redirects budget toward racks~B and~C, where it translates into higher satisfied demand.

The condition for inferior performance of the Greedy algorithm is a mismatch between aggregate demand and deliverable capacity within a subtree.
This can occur when:
(i)~server or rack capacities are heterogeneous (e.g., mixed hardware generations sharing the same PDN);
(ii)~workload placement concentrates high-power jobs under a subset of capacity-limited servers; or
(iii)~aggressive oversubscription at one level creates a tight bottleneck that Greedy algorithm's local split cannot anticipate.
Whenever such a mismatch exists, a global optimizer like nvPAX can redirect power where it is actually deliverable.

\section{nvPAX with Tenant SLA Constraints}
\label{app:sla-experiment}


The  experiment in Section~\ref{sec:numerical-results} compares nvPAX against Static and Greedy allocations under device limits and hierarchical PDN constraints only, without tenant SLA constraints.
This appendix demonstrates nvPAX's ability to enforce horizontal tenant SLA constraints -- minimum and maximum aggregate power guarantees per tenant -- while maintaining high device satisfaction.

\subsection{Experimental Setup}

\paragraph{Tenant and SLA Generation.}
We generate 100 synthetic tenants, each assigned 100~GPUs (fixed throughout the test).
The SLA bounds are set to 40\%--80\% of the tenant's maximum aggregate power: with 100~GPUs at 700~W each, this yields $B_k^{\min} = 28{,}000$~W and $B_k^{\max} = 56{,}000$~W per tenant.
These bounds represent a contractual guarantee that each tenant receives at least 40\% of its potential capacity while not exceeding 80\%, leaving headroom for other tenants under shared PDN constraints.

\paragraph{Device Priorities.}
Devices belonging to tenants are assigned random priorities $p_i \in \{1, 2, 3\}$, where higher values indicate higher priority.
Phase~I of nvPAX processes devices from highest to lowest priority, ensuring that higher-priority workloads receive allocations closer to their requests before lower-priority ones.
Devices not assigned to any tenant retain the default priority of~1.

\paragraph{Data and Simulation.}
We use the same production datacenter GPU telemetry as in Section~\ref{sec:numerical-results}.
The simulation runs for 3~days (8{,}523 timestamps at 30~seconds spacing) over more than 12,000~GPUs across 4~halls.
Device limits are $l_i = 200$~W and $u_i = 700$~W (H100 specifications), with oversubscription factor 0.85 and idle threshold of 150~W, identical to Section~\ref{sec:numerical-results}.

\subsection{SLA-Specific Metrics}

In addition to the global satisfaction ratio $S$ defined in Section~\ref{sec:numerical-results}, we report tenant-level metrics.
Let~$\mathcal{T}_k$ denote the set of devices assigned to tenant $k$, and let $B_k^{\min}$, $B_k^{\max}$ be the tenant's SLA bounds.

\begin{enumerate}
\item \textbf{Per-tenant satisfaction ratio:}
\[
S_k = \frac{\sum_{i \in \mathcal{T}_k} \min(r_i, a_i)}{\sum_{i \in \mathcal{T}_k} r_i}
\]
is the fraction of tenant $k$'s aggregate demand that is satisfied.

\item \textbf{SLA margin (lower bound headroom):}
\[
M_k^{\min} = \frac{\sum_{i \in \mathcal{T}_k} a_i - B_k^{\min}}{B_k^{\max} - B_k^{\min}}.
\]
A value $M_k^{\min} \geq 0$ indicates that the minimum SLA constraint is satisfied; larger values indicate greater headroom above the guaranteed minimum.

\end{enumerate}

\subsection{Results}

For \textbf{global satisfaction ratio}: nvPAX achieved mean 98.93\% (std 0.52\%, min 96.34\%, max 100.00\%).

For \textbf{per-tenant satisfaction}: mean across all tenants and timestamps was 99.24\% (std 0.37\%, min 97.46\%, max 100.00\%).

For \textbf{SLA margin}: at each timestamp, we first compute the mean $M_k^{\min}$ across all 100 tenants. We then average this per-timestamp mean over all 8{,}523 timestamps, yielding 54.44\% (std 0.63\%, min 44.16\%, max 59.99\%).
To evaluate the worst-case tenant, at each timestamp we find the minimum $M_k^{\min}$ across all 100 tenants. Averaging this per-timestamp minimum over all 8{,}523 timestamps yields 33.80\% (min 21.70\%, max 51.44\%), indicating that even the most constrained tenant at any given time maintained substantial headroom above its minimum guarantee.

During the test,  all timestamps  had all 100 tenants satisfy their SLA constraints.
There were zero minimum SLA violations and zero maximum SLA violations across all timestamps and tenants.

\textbf{Runtime} of nvPAX with SLA constraints (single \texttt{optimize()} call per timestamp): mean 718.83~ms, std 97.76~ms, min 535.45~ms, max 1{,}863.52~ms over 8{,}523 timestamps.
This is approximately 1.7 times slower than the non-SLA experiment (264.69~ms mean) due to the additional 200 tenant constraints (one minimum and one maximum bound per tenant).

\subsection{Discussion}

The results demonstrate that nvPAX successfully enforces horizontal tenant SLA constraints while maintaining high device satisfaction.
In this generated tenant experiment, the global satisfaction ratio with SLA constraints (98.93\%) is essentially the same as the non-SLA  experiment in Section~\ref{sec:numerical-results} (98.92\%).
Thus, for this trace and tenant construction, the additional horizontal constraints did not  reduce aggregate request satisfaction, while they did enforce contractual tenant-level bounds that the Static and Greedy baselines cannot represent.

The zero minimum-SLA and maximum-SLA violations across all 8{,}523 timestamps validate the constrained-optimization formulation for allocation under tenant SLA constraints.
Moreover, the positive lower-SLA margin indicates robustness beyond bare compliance: the mean tenant lower-SLA margin is 54.44\%, and even the worst tenant at each timestamp maintains a positive average margin of 33.80\% above the contractual minimum.
The minimum observed worst-tenant margin remains positive (21.70\%), showing that the allocation did not merely satisfy SLAs at numerical tolerance but maintained nontrivial headroom throughout the run.

The cost of this additional expressiveness is runtime: the SLA-constrained run takes 718.83~ms per timestamp on average, compared with 264.69~ms for the non-SLA run.
This overhead is expected because the optimizer must enforce 200 additional tenant constraints (one lower and one upper aggregate bound per tenant) while also processing nontrivial job priorities.